\documentclass[%
reprint,
superscriptaddress,
amsmath,amssymb,
aps,
prb,
]{revtex4-2}

\usepackage{lineno}
\usepackage{graphicx}
\usepackage{dcolumn}
\usepackage{bm}
\usepackage{subfigure}
\usepackage{float}
\usepackage{upgreek}

\begin{document}

	\preprint{APS/123-QED}
	\title{Stark tuning of telecom single-photon emitters based on a single Er$^{3+}$}
	
	\author{Jian-Yin Huang}
	\affiliation{CAS Key Laboratory of Quantum Information, University of Science and Technology of China, Hefei, 230026, China}
	
	\affiliation{%
		CAS Center for Excellence in Quantum Information and Quantum Physics, University of Science and Technology of China, Hefei, 230026, China 
	}%

    \author{Peng-Jun Liang}
    \affiliation{CAS Key Laboratory of Quantum Information, University of Science and Technology of China, Hefei, 230026, China}
    
    \affiliation{%
    	CAS Center for Excellence in Quantum Information and Quantum Physics, University of Science and Technology of China, Hefei, 230026, China 
    }%

    \author{Liang Zheng}
    \affiliation{CAS Key Laboratory of Quantum Information, University of Science and Technology of China, Hefei, 230026, China}
    
    \affiliation{%
    	CAS Center for Excellence in Quantum Information and Quantum Physics, University of Science and Technology of China, Hefei, 230026, China 
    }%

	\author{Pei-Yun Li}
	\affiliation{CAS Key Laboratory of Quantum Information, University of Science and Technology of China, Hefei, 230026, China}
	
	\affiliation{%
		CAS Center for Excellence in Quantum Information and Quantum Physics, University of Science and Technology of China, Hefei, 230026, China 
	}%

	\author{You-Zhi Ma}
	\affiliation{CAS Key Laboratory of Quantum Information, University of Science and Technology of China, Hefei, 230026, China}
	
	\affiliation{%
		CAS Center for Excellence in Quantum Information and Quantum Physics, University of Science and Technology of China, Hefei, 230026, China 
	}%

	\author{Duan-Chen Liu}
	\affiliation{CAS Key Laboratory of Quantum Information, University of Science and Technology of China, Hefei, 230026, China}
	
	\affiliation{%
		CAS Center for Excellence in Quantum Information and Quantum Physics, University of Science and Technology of China, Hefei, 230026, China 
	}%

	\author{Jing-Hui Xie}
	\affiliation{CAS Key Laboratory of Quantum Information, University of Science and Technology of China, Hefei, 230026, China}
	
	\affiliation{%
		CAS Center for Excellence in Quantum Information and Quantum Physics, University of Science and Technology of China, Hefei, 230026, China 
	}%

	\author{Zong-Quan Zhou} \thanks{email:zq\_zhou@ustc.edu.cn}
	\affiliation{CAS Key Laboratory of Quantum Information, University of Science and Technology of China, Hefei, 230026, China}
	
	\affiliation{%
		CAS Center for Excellence in Quantum Information and Quantum Physics, University of Science and Technology of China, Hefei, 230026, China 
	}%
	
	\affiliation{Hefei National Laboratory, University of Science and Technology of China, Hefei 230088, China}%
	\author{Chuan-Feng Li} \thanks{email:cﬂi@ustc.edu.cn
	}
	\affiliation{CAS Key Laboratory of Quantum Information, University of Science and Technology of China, Hefei, 230026, China}
	
	\affiliation{%
		CAS Center for Excellence in Quantum Information and Quantum Physics, University of Science and Technology of China, Hefei, 230026, China 
	}%
	
	\affiliation{Hefei National Laboratory, University of Science and Technology of China, Hefei 230088, China}%
	\author{Guang-Can Guo} 
	
	\affiliation{CAS Key Laboratory of Quantum Information, University of Science and Technology of China, Hefei, 230026, China}
	
	\affiliation{%
		CAS Center for Excellence in Quantum Information and Quantum Physics, University of Science and Technology of China, Hefei, 230026, China 
	}%
	
	\affiliation{Hefei National Laboratory, University of Science and Technology of China, Hefei 230088, China}%

	\date{\today}

	\begin{abstract}

  The implementation of scalable quantum networks requires photons at the telecom band and long-lived spin coherence. The single Er$^{3+}$ in solid-state hosts is an important candidate that fulfills these critical requirements simultaneously. However, to entangle distant Er$^{3+}$ ions through photonic connections, the emission frequency of individual Er$^{3+}$ in solid-state matrix must be the same, which is challenging because the emission frequency of Er$^{3+}$ depends on its local environment. Herein, we propose and experimentally demonstrate the Stark tuning of the emission frequency of a single Er$^{3+}$ in a Y$_2$SiO$_5$ crystal by employing electrodes interfaced with a silicon photonic crystal cavity. We obtain a Stark shift of 182.9 $\pm$ 0.8 MHz which is approximately 27 times of the optical emission linewidth, demonstrating the promising applications in tuning the emission frequency of independent Er$^{3+}$ into the same spectral channels. Our results provide a useful solution for construction of scalable quantum networks based on single Er$^{3+}$ and a universal tool for tuning emission of individual rare-earth ions.
  \\\\
\noindent{PACS: 03.67.Hk, 42.50.-p }

	\end{abstract}

	\maketitle
	The photon loss is exponentially dependent on the length of fiber channels which prevent the long-distance transmission of quantum information. The quantum repeater approach has been proposed to solve this problem \cite{briegel1998quantum, sangouard2011quantum} and has been implemented with various atomic systems \cite{ bernien2013heralded, moehring2007entanglement, chou2007functional, liu2021heralded, hensen2015loophole, yu2020entanglement,yuan2008experimental,hofmann2012heralded,delteil2016generation}. These demonstrations are limited to a short distance due to the fact that the optical transitions of these atomic systems are far away from the telecom band. Although non-degenerate photon pairs \cite{simon2007quantum,fekete2013ultranarrow,lago2021telecom, rakonjac2021entanglement} and quantum frequency conversion \cite{zaske2012visible,ikuta2011wide,radnaev2010quantum} have been employed in several works, they could introduce additional losses and noise.
	
	Since Er$^{3+}$ has natural optical transition in the telecom C band with long optical \cite{bottger2009effects} and spin coherence lifetimes \cite{ranvcic2018coherence, huang2022extending, rakonjac2020long}, it has been extensively studied in quantum networking applications, both as absorptive ensemble-based memories \cite{PhysRevLett.104.080502,saglamyurek2015quantum,askarani2019storage,craiciu2019nanophotonic,stuart2021initialization,craiciu2021multifunctional,PhysRevLett.129.210501} and as emissive single-atom-based emitters \cite{dibos2018atomic,raha2020optical,chen2020parallel,uysal2023coherent,ourari2023indistinguishable,ulanowski2022spectral}. By coupling with high-$Q$ micro/nano-photonic cavities, photoluminescence of Er$^{3+}$ is enhanced in orders of magnitude and single Er$^{3+}$ ions are able to be addressed optically \cite{dibos2018atomic,ulanowski2022spectral}. Spin manipulation and single-shot readout are subsequently demonstrated \cite{raha2020optical,chen2020parallel}, as well as controlling nearby nuclear spin as ancilla qubits \cite{uysal2023coherent}. Recently, the indistinguishability of successively emitted photons from a single Er$^{3+}$ ion  is verified with Hong-Ou-Mandel (HOM) interference \cite{ourari2023indistinguishable}. For a single Er$^{3+}$ quantum light source to be applicable in a realistic quantum repeater network, indistinguishablility should be guaranteed for photons emitted from independent Er$^{3+}$ ions as a prerequisite to achieve entanglement swapping. However, because of complicated local environments inside a solid, different Er$^{3+}$ dopants exhibit different optical transition frequencies. This could enable the spectral addressing of many independent ions \cite{chen2020parallel}, but the problem of wavelength alignments must be solved. For Er$^{3+}$ ions coupled with nanophotonic cavities, the inhomogeneous broadening can typically reach several GHz, while the ions with sufficient Purcell enhancement are rare \cite{dibos2018atomic}. Therefore, a large-scale spectral tuning greater than 100 MHz is profitable to achieve indistinguishability.  In this Letter, we propose and demonstrate Stark tuning of Er$^{3+}$ in Y$_2$SiO$_5$, which has shown to be a simple and robust tool to tune spectra of single-atom based emitters. 
	
	Y$_2$SiO$_5$ is an often used host material due to its small magnetic moments (-0.137 $\mu_N$ for ${}^{89}$Y) and low natural abundance of magnetic isotopes (4.7\% with -0.554 $\mu$$_N$ for ${}^{29}$Si, 0.04\% with -1.89 $\mu$$_N$ for ${}^{17}$O ). In this work, the experiments were performed based on Y$_2$SiO$_5$, which have displayed good coherent properties for Er$^{3+}$ \cite{bottger2009effects, ranvcic2018coherence,huang2022extending, rakonjac2020long}. The Y$_2$SiO$_5$ crystal was grown by the Czochralski method in a 35-kW inductively heated generator at a 3-kHz middle frequency. The crystal is then cut along the  optical extinction axes of D$_1$, D$_2$, and $b$ with dimensions of $10\times 8\times 3$ mm$^3$. The purity of raw metarials used for crystal growth is at least 99.999\%. The concentration of Er$^{3+}$ is less than 0.5 ppm as verified by the glow discharge mass spectrometry.

 \begin{figure*}
   	\centering
   	\subfigure{
   		\includegraphics[width=2\columnwidth]{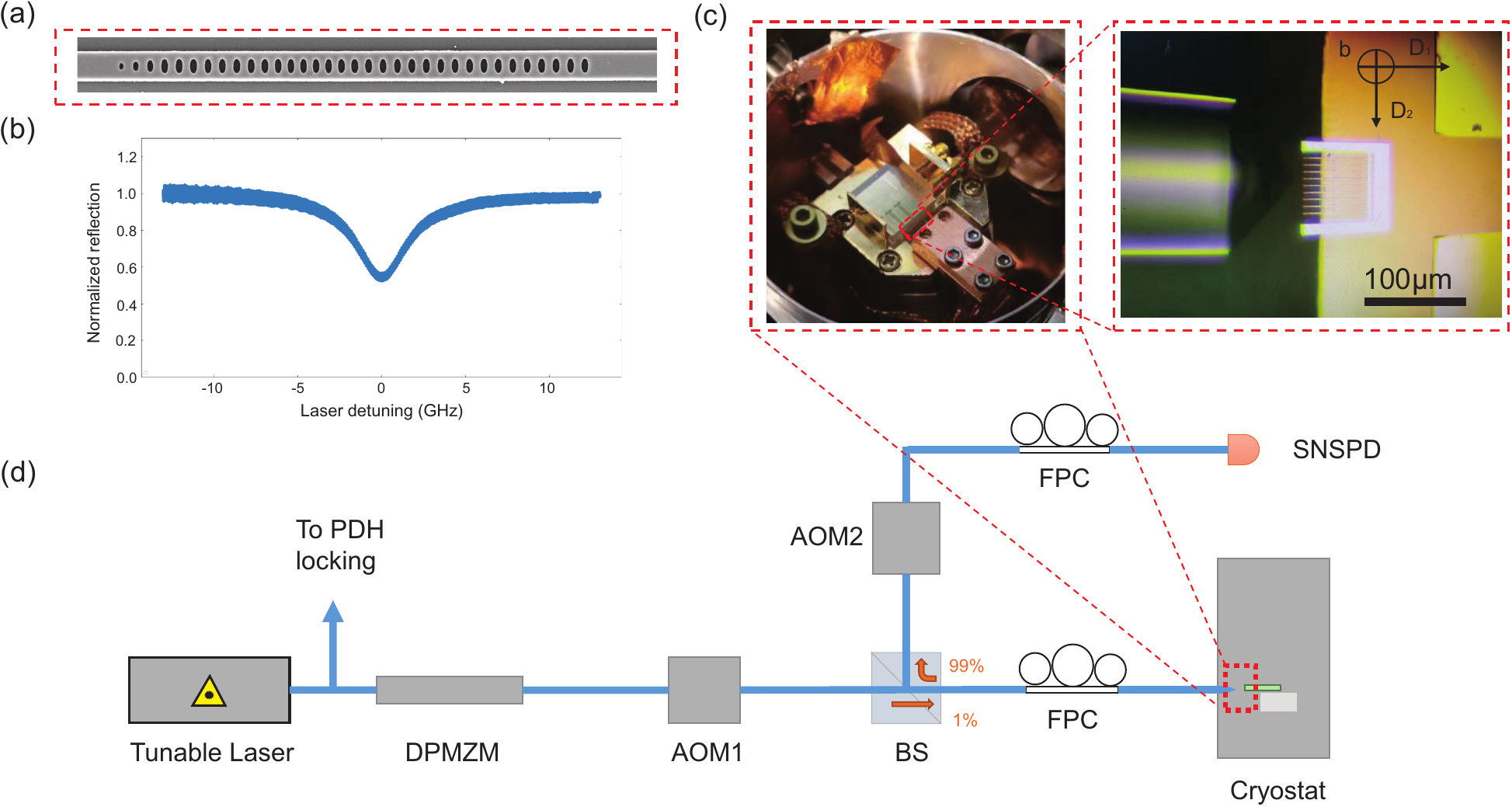}}
   	
   	\caption{(a) The scanning electron microscope image of the silicon photonic cavity structure before transferring to the surface of Y$_2$SiO$_5$.  (b) Frequency reflection spectrum of the silicon photonic cavity used in these experiments, with a quality factor of  5.1$\times$10$^4$. (c) Top-view image of the sample inside the cryostat. The photonic crystal cavity is placed on the D$_1$D$_2$ plane of the Y$_2$SiO$_5$ surface and in front of the electrode. The part that emerges from the edge of the sample is tapered waveguides, and they can be coupled to the lensed fiber with edge coupling. (d) Simplified diagram of the experimental setup. The laser frequency is stabilized to a high-finesse optical cavity . A dual-parallel Mach-Zehnder modulator (DPMZM) is employed for wideband sweeping of the excitation frequency during the acquisition of PLE spectra. The continuous laser is chopped by AOM1 into pulses to excite the ions. AOM2 is used to block the excitation light from entering the superconducting nanowire single-photon detector (SNSPD). The fiber polarization control (FPC) is used to control the laser polarization. The sample is situated in a 4 K cryostat. } 
   \end{figure*}

 Single Er$^{3+}$ ions need to be identified before studying the Stark effect. Since the optical excited state lifetime of Er$^{3+}$ in bulk Y$_2$SiO$_5$ is as long as 11 ms \cite{bottger2006spectroscopy}, the photon emission from a
single Er$^{3+}$ is too weak to be detected. Here, we follow the method of ref. \cite{dibos2018atomic} which adopted photonic crystal (PC) cavities to interface with Er$^{3+}$ ions for obtaining Purcell enhancement on the photon emission rate, and to improve the photon collection efficiency. The PC is fabricated on a silicon on insulator (SOI) wafer with a 220 nm device layer by electron beam lithography (EBL) and inductively coupled plasma (ICP), and then transferred onto the Y$_2$SiO$_5$ crystal via the stamping technique \cite{lee2014heterostructures}.  A scanning electron microscope image (SEM) of the PC is presented in Fig. 1a.  The quality factor of the PC cavity is approximately 5.1$\times$10$^4$ as measured with the cavity reflection spectrum (Fig. 1b). In order to achieve Stark tuning of the emission frequency, two additional electrodes are patterned on the host crystal's surface near the PC cavity to apply the required electric field on the target Er$^{3+}$ ions. The electrode structure is fabricated with ultraviolet lithography, electron beam evaporation, and a lift-off process. The electrodes are 200 $\upmu$m wide and are 100 $\upmu$m apart from each other. The layout of the PC, the Y$_2$SiO$_5$ crystal and the electrodes can be found in Fig. 1c. During the fabrication of the electrodes, the photoresist tends to be thicker at the crystal's edge compared to that at the center. Meanwhile, the Y$_2$SiO$_5$ crystal is too thick and too brittle to allow chip-free splicing. For a lift-off process with sufficient quality, two types of photoresists are spin-coated into two layers (LOR5A followed with S1813, both with 3000 rpm and 30 s), and a long exposure time of 51 s is adopted. 
	
Fig. 1d presents the experimental configuration. A small portion of the laser (Toptica CTL 1500) is guided to a reference cavity to stabilize the laser frequency with the Pound–Drever–Hall (PDH) technique \cite{black2001introduction}. To scan the excitation frequency continuously, most laser power passes through a dual-parallel Mach-Zehnder modulator for single sideband frequency modulation. The laser pulses are chopped by using two consecutive double-passed acousto-optic modulators (AOMs), then enters the PC cavity via edge coupling to excite the ions. The fluorescence is gathered to the superconducting single-photon detector through the 99:1 beam splitter (BS) and two consecutive AOMs, which are used to prevent excitation pulses from influencing the detector. Two fiber polarization controllers (FPCs) are used to match the polarization to the PC cavity and to the detector. The single-photon detector has an 80\% detection efficiency and 2 Hz dark counts. The overall photon collection efficiency is 1\%, including excitation efficiency, transmission efficiency, and detector efficiency.

     \begin{figure*}
    	\subfigure{
    		\includegraphics[width=2.1\columnwidth]{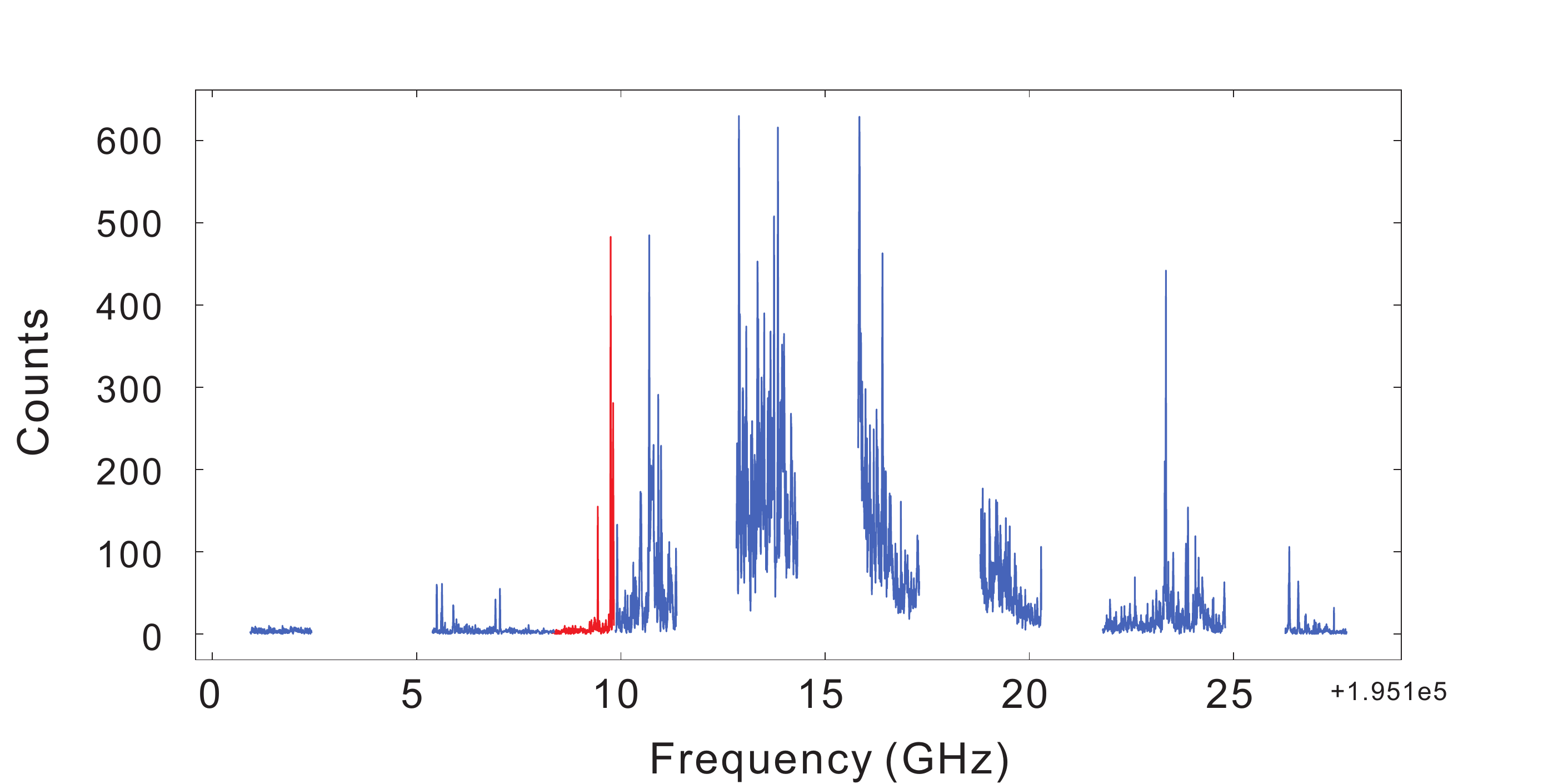}}
    	
    	\caption{PLE spectrum for norminally undoped Y$_2$SiO$_5$ crystal. The excitation frequency is swept around a center of 195115 GHz. The studied frequency range covers the absorption band of ensemble Er$^{3+}$ in Y$_2$SiO$_5$. The intervals in the spectrum are caused by technical limitations in the vacuum degree of the current cryostat \cite{dibos2018atomic}. During the acquisition of the PLE spectrum, 10 $\upmu$s excitation pulses are repeated at a frequency of 10 kHz. Detection windows are arranged 1 $\upmu$s after the excitation pulses with a length of 85 $\upmu$s, during which AOM2 is turned on. The integration time is 5 s for each frequency point. The scan pitch is set as 5 MHz. } 
    \end{figure*}

    \begin{figure*}
    	\centering
    	\subfigure{
    		\includegraphics[width=2.1\columnwidth]{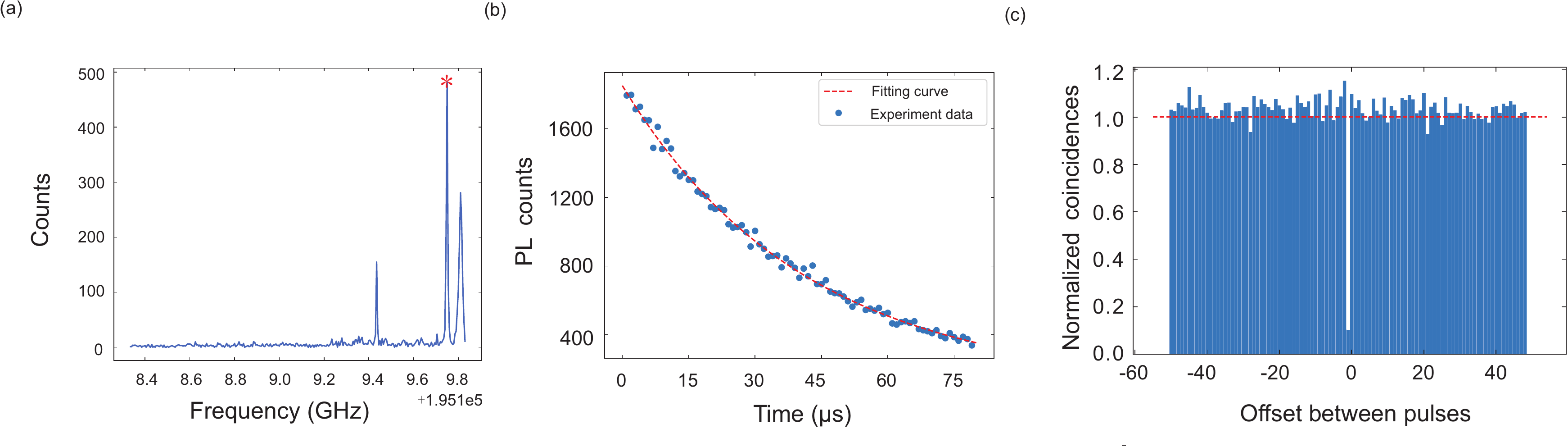}}
    	\caption{(a) Magnified image of the red-line area in Fig. 2.  (b)  Fluorescence decay of a cavity-coupled single Er$^{3+}$ ion. Experimental data are given in blue dots; the exponentially fitted curve is shown by a red dashed line. (c) Second-order autocorrelation function g$^{(2)}$ measured for single Er$^{3+}$ ion.
        The data in (b) and (c) were measured at the peak marked by asterisk in the PLE spectrum given in (a). 
    	 }
    \end{figure*}

We searched for single Er$^{3+}$ by photoluminescence excitation (PLE) spectroscopy  using resonant excitation. The excitation window and the detection window was separated in time. The PLE spectrum is given in Fig. 2. The series of sharp peaks are considered to result from photons emitted by individual Er$^{3+}$ ions. Fig. 3a is a magnified image of the red-line area in Fig. 2. The peak marked with an asterisk is a representative single-ion PLE spectrum. The optical fluorescence lifetime of this ion was measured to be 41.0 $\pm$ 1.4 $\upmu$s, as shown in Fig. 3b. This has been 278 times shorter compared with the bulk lifetime of 11.4 ms \cite{bottger2006spectroscopy}, showing the Purcell enhancement enabled by the atom-cavity interaction.  The autocorrelation function g$^{(2)}$(0) \cite{reiserer2015cavity} is measured to be 0.10 $\pm$ 0.04 for this emission peak (Fig. 3c), confirming that this peak is indeed the signal from a single Er$^{3+}$ ion. Here we denote this ion as ``ion 1" for convenience in the following discussion. Although there are more peaks in the center of the inhomogeneous line, these peaks are typically overlayed with strong background noise coming from weakly-coupled ions, which would degrade the quality of output photons. Thus, all the ions in this study are searched on the tail of the inhomogeneous distribution.

   The frequency shift of the atomic optical transition  caused by a static electric field is well known as the DC Stark effect, which is the result of the differences in dipole moments and the polarizability between the excited and ground states. Usually, the induced frequency shift can be expressed as \cite{macfarlane2007optical}:
   
   \begin{equation}
    h\Delta \nu = -\Delta \bm{\mu} \cdot \textbf{L} \cdot \textbf{E} -\frac{1}{2}  \textbf{E} \cdot \textbf{L}  \cdot \Delta \bm{\alpha} \cdot \textbf{L} \cdot \textbf{E}+\cdots
   \end{equation}
    where $h$ is the Planck constant; $\Delta \nu$ represents the optical frequency shift; $\textbf{L}$ is the local field correction tensor; $\textbf{E}$ is the electric field. $\Delta \mu$ and $\Delta \alpha$ are the differences between the ground and excited states in permanent dipole moment and polarizability, respectively. The omitted terms depend on the product of the local electric field with  higher order hyperpolarization. The first-order linear Stark shift depends on the expectation values of the electric dipole moments for the optical ground and excited states. These values would vanish for the 1.5-$\upmu$m 4f-4f transition of free Er$^{3+}$ ions, or Er$^{3+}$ ions occupying crystallographic sites with non-polar site symmetry. When doping into Y$_2$SiO$_5$, trivalent rare-earth ions such as Er$^{3+}$ usually replace Y$^{3+}$ which locate at sites with very low symmetry of C$_1$, and the linear Stark effect is maintained. In previous ensemble-based studies on rare-earth-doped Y$_2$SiO$_5$ crystals, DC Stark shifts are dominated by the linear terms, and higher-order effects can be omitted \cite{horvath2021noise,liu2020demand,craiciu2021multifunctional,PhysRevLett.129.210501}. According to the measured data presented in Fig. 4, the Stark shifts of single Er$^{3+}$ are also dominated by the linear terms, which are in consistence with the ensemble measurements, although much larger electric fields are applied compared with previous works.

   \begin{figure}
   	
   	\subfigure{
   		\includegraphics[width=1\columnwidth]{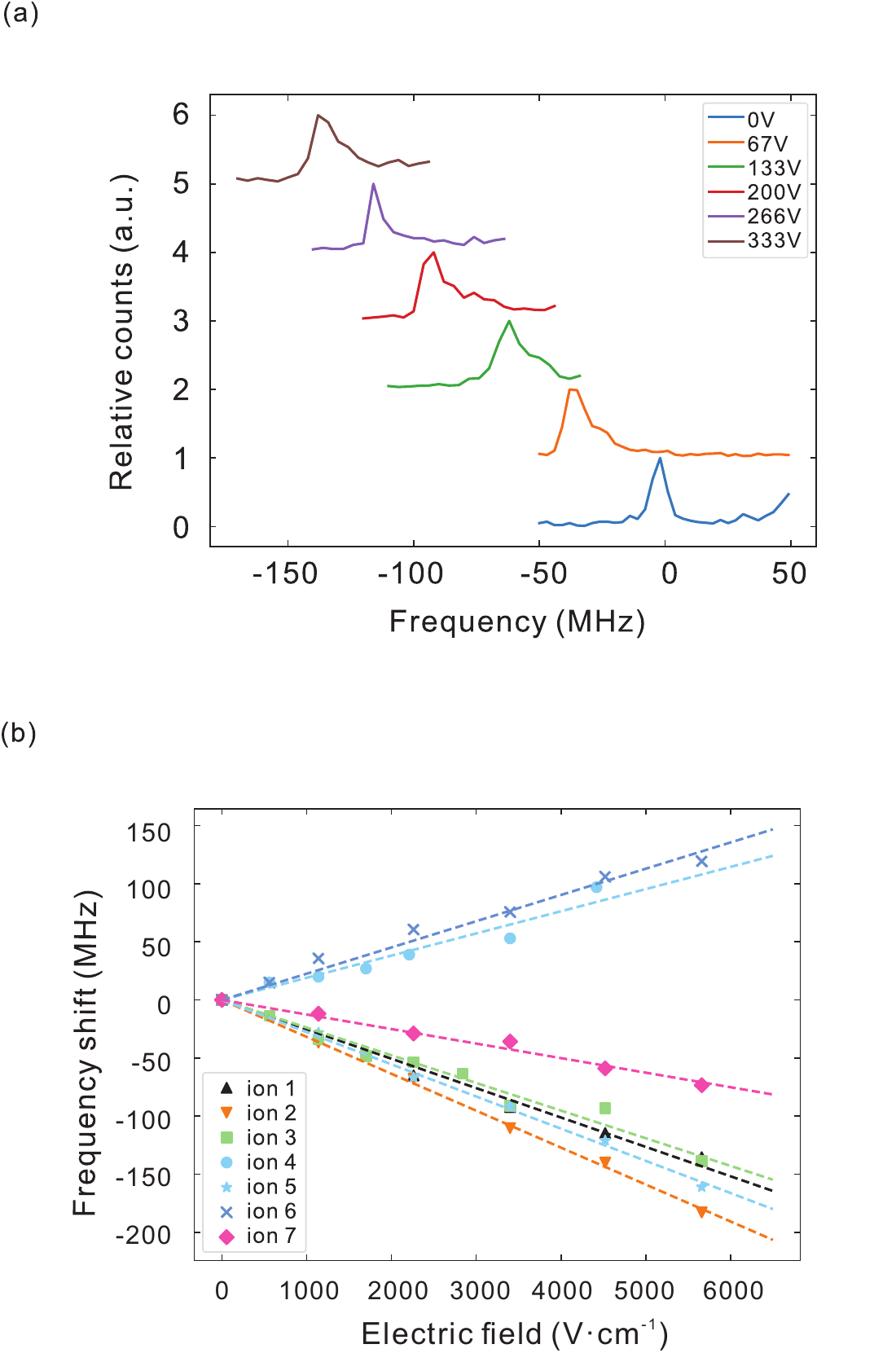}}\\

   	
   	\caption{(a) Frequency shifts of ion 1 as response to different voltages applied to the electrodes. (b) Stark shifts of multiple individual Er$^{3+}$ ions. Measured data for each ion are linearly fitted.  }
   \end{figure}

   In this work, the external electric field generated by the electrodes is aligned along the crystal's D$_2$ axis. A series of voltages are applied to the electrodes, and the frequency shifts of individual Er$^{3+}$ ions are measured through PLE scanning. PLE spectra of ion 1 corresponding to different applied voltages are presented in Fig. 4a. It can be observed that the optical transition frequency shifts linearly along with the increased voltages. Since both $\Delta \bm{\mu}$ and  $\textbf{L}$ in Eq.(1) are anisotropic and unknown, here the linear Stark effect is described empirically as $\Delta \nu $ = $s_{\hat{n}} E_{\hat{n}}$, where $\hat{n}$ represents the unit vector along the D$_2$ axis; $E_{\hat{n}}$ is the electric field strength, and $s_{\hat{n}}$ is the Stark coefficient. Due to technological limitations, the edge of the crystal is not suitable for gilding. As a result the PC is not in the center of the electrode, which can be seen in Fig. 1(c). Considering that the mode volume of the PC cavity is small, we approximate that the electric field at the center of the PC cavity is 
   the electric field that the ion is experiencing.  The values of $E_{\hat{n}}$ are then extracted with numerical simulation based on the input voltages. $s_{\hat{n}}$ of ion 1 is fitted as 19.8 $\pm$ 0.5 kHz/(V$\cdot$cm$^{-1}$). The fitted line is given in Fig. 4b. Fig. 4a also reveals that the linewidth of the single Er$^{3+}$ ion is slightly broadened as the voltage increases. This may be due to the fact that the increased voltage has made the Er$^{3+}$ more sensitive to charge noise in its surroundings, which is similar to the phenomenon observed in color centers in diamonds \cite{de2021investigation,tamarat2006stark}. During the measurements the resonance frequency of the nanocavity experienced a persistent redshift of approximately 3 GHz/h, probably due to gas condensation \cite{dibos2018atomic}. To finish each PLE scanning in time, the corresponded frequency range of each curves in Fig. 4(a) is relatively narrow. Nevertheless, as shown in Fig. 3(a), the next peak in the redshift region is more than 300 MHz away from  emission of ion 1, while the maximum tuning range for ion 1 is only 150 MHz. In addition, the frequency shift of these peaks has a linear dependence on the applied voltage. Therefore, these peaks displayed in Fig. 4(a) should all originate from ion 1. To recover the cavity resonance frequency for obtaining longer times for data acquisition, continuous laser light with a power greater than 100 $\upmu$W can be sent into the cryostat to sublimate the condensed gases on the nanocavity.
 
   The Stark shifts of multiple individual Er$^{3+}$ ions were further measured, and the results are presented in Fig. 4b. There appears to be two classes of ions experiencing red and blue shifts, respectively. Under an arbitrary external electric field, there can be up to four possibilities of Stark shifts for Er$^{3+}$:Y$_{2}$SiO$_{5}$ because of the four orientations of a single crystallographic site \cite{guillot2006hyperfine}. Here $\textbf{E}$ is perpendicular to the b axis, and the four Stark shifts are pairwise degenerate. This leads to two types of Stark shifts with the same magnitude while in opposite directions. As shown in Fig. 4b,  The Stark coefficients for different Er$^{3+}$ ions are found to be diverse. For ions 1 to 6, $\lvert s_{\hat{n}}\rvert$ has an average value of 20 kHz/(V$\cdot$cm$^{-1}$) and a standard deviation of 5.8 kHz/(V$\cdot$cm$^{-1}$). The average value is consistent with the values reported in the ensemble-based experiment \cite{minavr2009electric}. However, the deviation is unexpectedly large. In previous works \cite{craiciu2021multifunctional,PhysRevLett.129.210501},  which are also ensemble-based experiments working on the same material while $\textbf{E}$ is aligned parallel to the b axis, only slight antihole broadening is observed when DC voltage is applied. We speculate that this is because the Er$^{3+}$ ions under study in this work are the ions with relatively large Purcell effects. These ions are very close to the surface and their local environments are prone to be more complicated.
   We have also found one ion (ion 7) with an exceptionally deviated $\lvert s_{\hat{n}}\rvert$ of $9.8\pm0.3$ kHz/(V$\cdot$cm$^{-1}$), indicating that a significant defect may exist near this ion and such Stark-measurements could be a useful tool in probing local environments. Under the highest applied voltage of 333 V, the largest frequency shift of 182.9 $\pm$ 0.8 MHz is achieved for ion 2, which is 27.3 $\pm$ 1.2 times compared with its zero-field linewidth of 6.7 $\pm$ 0.3 MHz.

   In conclusion, we have performed a study on the Stark shift of single Er$^{3+}$ ions in Y$_2$SiO$_5$. The Stark effect can be studied more
   directly and refined using the single-ion method compared with the ensemble method. In this work the maximum frequency shift reaches 182.9 $\pm$ 0.8 MHz, which has been sufficient to tune two independent Er$^{3+}$ ions into resonance. Greater frequency shifts can be achieved by further increasing the DC voltage or reducing the distance between the electrodes. For example, by further shrinking the electrode separation by 5 times and to mildly increase the applied voltages, Stark shift greater than 0.5 GHz can be expected. Meanwhile, the risk of possible dielectric breakdown caused by large electric field should be noticed and more tests should be performed. Our results demonstrate that Stark tuning is an efficient method for adjusting the emission wavelength of individual Er$^{3+}$ ions, and it can be useful in future efforts to generate entanglement between two distant Er$^{3+}$ ions in a quantum network. 

   J.-Y. H. and P.-J. L. contributed equally to this work.
  \\
  \\
  {\it Note Added.}
   Recently, we became aware of independent work on the Start tuning of single Er$^{3+}$ emission in the LiNbO$_3$ crystal by Yong Yu et al \cite{yu2023frequency}.
   
   	\begin{acknowledgments}
   	This work was supported by the National Key R\&D Program of China (No. 2017YFA0304100), the Innovation Program for Quantum Science and Technology (No. 2021ZD0301200), the National Natural Science Foundation of China (Nos. 12222411 and 11821404) and this work was partially carried out at the USTC Center for Micro and Nanoscale Research and Fabrication. Z.-Q. Zhou acknowledges the support from the Youth Innovation Promotion Association CAS. The authors thank W. Liu, L. Chen, Y.-Z. He, and C. Tan for their assistance in micro and nano fabrication.
   \end{acknowledgments}

	\nolinenumbers
	\bibliography{article}
	
\end{document}